\documentclass[aps,twocolumn,showpacs]{revtex4}
\usepackage{color}
\usepackage{graphicx}
\usepackage{amsmath}
\usepackage{amssymb}
\usepackage{amsfonts}
\usepackage[colorlinks,hyperindex]{hyperref}
\begin{document}
\title{Stability and spatial coherence of nonresonantly pumped exciton-polariton condensates}
\author{Nataliya Bobrovska$^1$, Elena A.~Ostrovskaya$^2$ and Micha{\l} Matuszewski$^1$}
\affiliation{
$^1$Institute of Physics Polish Academy of Sciences, Al. Lotnik\'ow 32/46, 02-668 Warsaw, Poland \\
$^2$Nonlinear Physics Centre, Research School of Physical Sciences and
Engineering, The Australian National University, Canberra ACT 0200,
Australia
}

\begin{abstract}
We investigate the stability and coherence properties of one-dimensional exciton-polariton condensates under
nonresonant pumping. We model the condensate dynamics using the open-dissipative Gross-Pitaevskii equation. 
In the case of spatially homogeneous pumping, we find that the instability of the steady state
leads to significant reduction of the coherence length. 
We consider two effects that can lead to the stabilization
of the steady state, i.e.~the polariton energy relaxation and the influence of an inhomogeneous pumping profile.
We find that, while the former has little effect on the stability, the latter is very effective in stabilizing the condensate
which results in a large coherence length.
\end{abstract}
%\keywords{exciton polariton condensate, stability analysis, generalized Gross-Pitaevskii equation (GPE)}
\maketitle

%===============================================================================
\section{Introduction}
%===============================================================================

Exciton-polaritons are light-matter bosonic quasiparticles created due to strong coupling between excitons 
and photons~\cite{kavokin2007-book,Weisbuch_Polaritons,Kasprzak_BEC}. Their extremely light effective mass combined with 
strong exciton-mediated
interparticle interactions makes them an ideal system for investigation of fundamental
phenomena such as room temperature Bose-Einstein condensation~\cite{Grandjean_RoomTempLasing,Kena_Organic,Mahrt_RTCondensatePolymer} 
as well as applications~\cite{Bramati_SpinSwitches,Savvidis_TransistorSwitch,Sanvitto_Transistor,
Hofling_ElectricallyPumped,Battacharya_RoomTemperatureElectrically}.
A wide range of phenomena observed in exciton-polariton systems, such as superfluidity~\cite{amo2009},
or spontaneous creation of self-localized structures, including solitons and vortices, 
has attracted great interest in recent years~\cite{Carusotto_QuantumFluids,Yamamoto_RMP}.
In contrast to condensates of ultracold atoms, polariton superfluids are nonequilibrium systems 
in which continuous pumping is required to maintain the condensate population~\cite{kavokin2007-book,Kasprzak_BEC,Yamamoto_RMP,wouters2007}.

In a number of experiments, spatial coherence extending over the whole polariton cloud 
was demonstrated~\cite{Kasprzak_BEC,Yamamoto_SpatialCoherence,Bloch_ExtendedCondensates,Hofling_1DcoherenceInteractions, Yamamoto_RMP}. 
Recently, stability limits for polariton condensates
under nonresonant pumping 
were determined within the open-dissipative Gross-Pitaevskii model~\cite{ostrovskaya2014}.
It was predicted that {\em spatially homogeneous} steady states are stable 
in regions of parameter space determined by the ratio $\gamma_R g_C / \gamma_C g_R$,
where $\gamma_{C,R}^{-1}$ are the lifetimes of the polariton condensate and the exciton reservoir
and $g_{C,R}$ are the coefficients of interaction within the condensate and between the condensate and the reservoir, respectively.
Importantly, for comparable interaction constants, stability close to the threshold 
was predicted only for relatively small values of $\gamma_C / \gamma_R$.
It is important to note that
values of  $\gamma_R$ varying by orders of magnitude are used throughout the literature~\cite{Bloch_1DAmplification,Deveaud_VortexDynamics,Yamamoto_VortexPair}.
According to independent measurements~\cite{Szczytko_ExcitonFormation,Bloch_DynamicsElectronGas}, 
physical values of $\gamma_R$ correspond to the unstable regime,
where, as we show below, a significant reduction of the coherence length can be expected.

In this work, we investigate whether large coherence length can emerge in condensates even in the unstable region of parameter
space predicted by the homogeneous theory.
We analyse the stability and coherence properties of polariton condesates in more detail,
taking into account the effects of polariton relaxation and the inhomogeneous pumping profile. 
We use the Bogoliubov-de Gennes theory as well as direct numerical integration of the open-dissipative Gross-Pitaeskii equations.
We demonstrate that full coherence can be achieved even for large values of $\gamma_C / \gamma_R$ once the finite size of the
pumping spot is taken into account. We believe that further experiments, in particular with 
homogeneous or ring-shaped pumping profiles, are necessary to determine stability limits in parameter
space and fix the values of phenomenological parameters of the model. In particular, the observation of coherence reduction
or its absence in the case of almost-homogeneous pumping would verify whether the ratio $\gamma_C / \gamma_R$ corresponds to the 
unstable regime.

%===============================================================================
\section{The model}
%===============================================================================

In one dimension (e.g.~in a microwire~\cite{Bloch_ExtendedCondensates}) the exciton-polariton condensate with the wavefunction  $\psi(x,t)$  
can be modelled with the generalized open-dissipative Gross-Pitaevskii equation coupled to 
the rate equation for the polariton reservoir density, $n_R(x,t)$ 
\cite{wouters2007,Xue_DarkSoliton}
%\begin{align}
\begin{equation}
\begin{split}
\label{GP1}
i \hbar\frac{\partial \psi}{\partial t} &=-\frac{\hbar^2 }{2 m^*} \frac{\partial^2 \psi}{\partial x^2} 
+ g_{\rm C}^{\rm 1D} |\psi|^2 \psi + g_{\rm R}^{\rm 1D} n_{\rm R} \psi \\
&+i\frac{\hbar}{2}\left(R^{\rm 1D} n_{\rm R} - \gamma_{\rm C} \right) \psi, \\
\frac{\partial n_{\rm R}}{\partial t} &= P(x) - (\gamma_{\rm R}+ R^{\rm 1D} |\psi|^2) n_{\rm R} 
\end{split}
\end{equation}
%\end{align}
 where $P(x)$ is the exciton creation rate determined by the pumping profile, $m^*$ 
is the effective mass of lower polaritons, 
$\gamma_{\rm C}$ and $\gamma_{\rm R}$ are the polariton and exciton loss rates, 
and $(R^{\rm 1D},g_i^{\rm 1D})=(R^{\rm 2D},g_i^{\rm 2D})/\sqrt{2\pi d^2}$ are
the rates of stimulated scattering into the condensate and the interaction coefficients, 
rescaled in the one-dimensional case. 
Here, we assumed a Gaussian transverse profile of $|\psi|^2$ and $n_{\rm R}$ of width $d$.
In the case of a one-dimensional microwire~\cite{Bloch_ExtendedCondensates}, 
the profile width $d$ is of the order of the microwire thickness.
We note that the exciton field correspond to the ``active'' exciton population rather 
than the reservoir at high energy 
levels~\cite{Deveaud_VortexDynamics}.
While the latter may have much longer lifetime $\gamma^{-1}$, it is not subject to a 
considerable back-action from polaritons, 
such as stimulated scattering, which is relevant for the stability properties of the system.

By rescaling time, space, wavefunction amplitude and material coefficients as $t=\tau \widetilde{t}$, $x=\xi\widetilde{x}$, $\psi=(\xi\beta)^{-1/2}\widetilde{\psi}$, $n_R=(\xi\beta)^{-1}\widetilde{n}_R$  $R^{\rm 1D}=(\xi\beta/\tau)\widetilde{R}$, $(g^{\rm 1D},g_R^{\rm 1D})=(\hbar\xi\beta/\tau)(\widetilde{g},\widetilde{g}_R)$, $(\gamma,\gamma_R)=\tau^{-1}(\widetilde{\gamma},\widetilde{\gamma}_R)$, $P(x)=(1/\xi\beta\tau)\widetilde{P}(x)$, where $\xi=\sqrt{\hbar \tau /2m^*}$, while $\tau$ and $\beta$ are arbitrary scaling parameters, we can rewrite the above equation in the dimensionless form (from now on we omit the tildas for convenience)
%\begin{flalign} 
\begin{align}
\label{psi}
&i\frac{\partial \psi}{\partial t}=\left[ -\frac{\partial^2}{{\partial x}^2}+\frac{i}{2}\left(Rn_R-\gamma\right)+g|\psi|^2+g_Rn_R\right]\psi,\\
\label{nr}
&\frac{\partial n_R}{\partial t}=P(x)-\left(\gamma_R+R|\psi|^2\right)n_R,
\end{align}
%\end{flalign}
%\end{eqnarray}
In particular, we may choose the time scaling $\tau$ in such a way that $\gamma_R=1$ without loss of generality. 
The norms of both fields $N_\psi=\int|\psi|^2 dx$ and $N_R=\int n_R dx$ are scaled by the factor of $\beta$.

%===============================================================================
\section{Homogeneous pumping}
%===============================================================================
We investigate the stability of a uniform condensate solution with the assumption of homogeneous pumping (${P=const}$)
%\begin{align}
\begin{equation}
\begin{split}
&\psi(x,t)=\psi_0 e^{-i\mu_0 t},\\
&n_R(x,t) = n_R^0,
\label{steady-state}
\end{split}
\end{equation}
where  $\mu_0$ is the chemical potential of the condensate. Substitution of the above equations into Eqs.~(\ref{psi}) gives the 
steady state solutions. For $P<P_{th}=\gamma\gamma_R/R$ only the noncondensed solution $\psi_0=0$ exists with  $n_R ^0=P/\gamma_R$. 
Above threshold, the condensate density is ${|\psi_0|^2=(P/\gamma)-(\gamma_R/R)}$ with the chemical potential $\mu_0=g|\psi_0|^2+g_Rn_R ^0$
and the reservoir density $n_R ^0=\gamma/R$ \cite{wouters2007,ostrovskaya2014}.

\begin{figure}
	\begin{center}
	\includegraphics[width=1.\columnwidth]{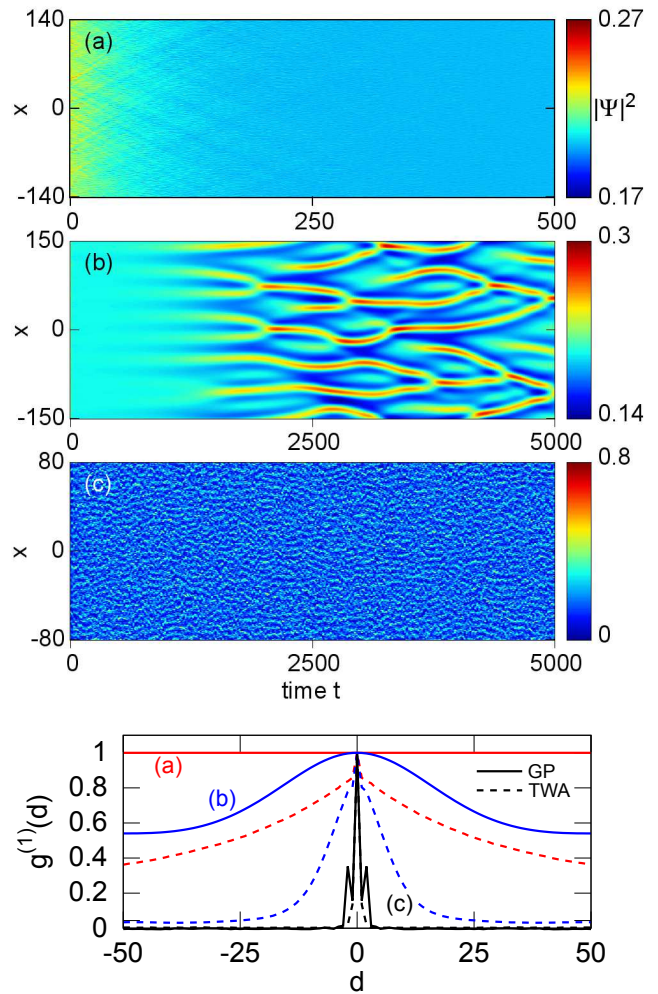} 
		\caption{(Colour online).~ Evolution of the stationary state ${\psi_0=\sqrt{(P-P_{th})/\gamma}}$ with additional noise in the case of homogeneous  pumping. The evolution of density in  (a) the stable case and (b), (c) unstable case near and far from the critical point, respectively. 
Parameters are $R=1$, $g=1$, $g_R=2g$, $P/P_{th}=1.2$ and $\gamma/\gamma_R=0.1, 0.66, 4.5$ for (a), (b), (c), respectively. Bottom panel shows the first-order correlation functions $g^{(1)}(d)=\langle \psi^*(x)\psi(x+d)
\rangle/\langle|\psi(x)|^2\rangle$ at ${t=t_{max}}$, calculated from about 500 samples.
Solid lines show $g^{(1)}$ calculated with Eqs.~(\ref{GP1}) and dashed lines depict the corresponding results with 
quantum fluctuations included (see text). 
Corresponding parameters in physical units are: time unit $\tau=\gamma_R^{-1}=10\,$ps, length unit $\xi=3.4\,\mu$m, $g=3.4\,\mu$eV$\mu$m$^2$, 
$R=5.1\times 10^{-3}\,\mu$m$^2$ ps$^{-1}$ for  $d=2\,\mu$m, $m^*=5\times 10^{-5} m_{\rm e}$, and $\beta=0.003$. 
TWA correlation lengths are (a) $\xi=70 \mu$m, (b) $\xi=17 \mu$m, and (c) $\xi=3 \mu$m.
}
		\label{fig:stab-noise}
	\end{center}
\end{figure}

\begin{figure}[tbp]
	\begin{center}
		\includegraphics[width=1.0\columnwidth]{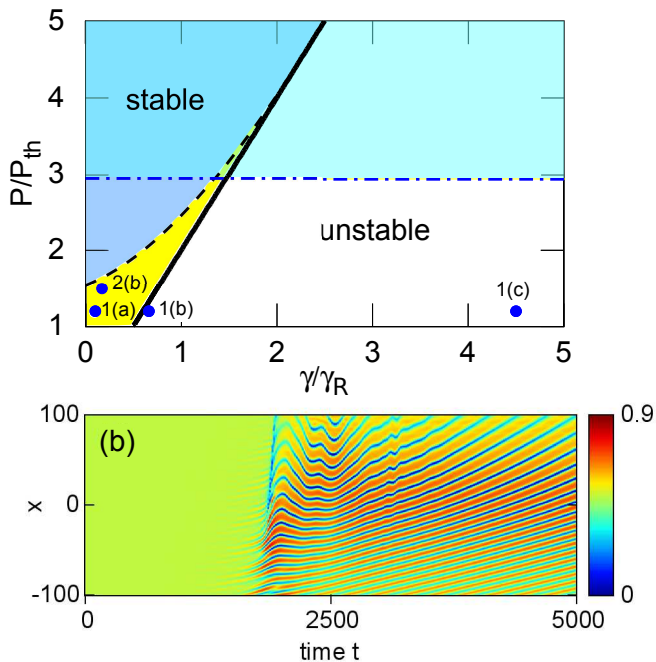} 
		\caption{(Colour online).~(a) Stability limits of a static condensate created with homogeneous pumping (solid line) 
and for a uniform condensate flowing with momentum $p=1$ (dashed line). Parameters are ${R=1}$, ${g=1}$, ${g_R=2g}$, ${\gamma_R=1}$. 
The dash-dotted line corresponds to the crtical velocity condition $p=c_s=\sqrt{2 g |\psi|^2}$.
Black solid line corresponds to the analytical condition~(\ref{condition}) for stability of the homogeneous solution. 
Blue points correspond to particular cases from Fig.~\ref{fig:stab-noise}. 
(b) Evolution of $|\psi(x,t)|^2$ for a uniform flow with homogeneous pumping and momentum $p=1$ in the motionally 
unstable regime, $P/P_{th}=1.5$, $\gamma/\gamma_R=0.17$.}
		\label{fig:with-p}
	\end{center}
\end{figure}
\par
The stability of the condensed state depends on the system parameters. In Fig.~\ref{fig:stab-noise} we show typical examples
of dynamics of the condensate density in stable, weakly unstable, and strongly unstable cases. 
Periodic boundary conditions are imposed which corresponds to a ring-shaped geometry of the microwire.
The initial state is perturbed by a white noise, which mimics classical (eg.~thermal) density fluctuations in the
initial polariton field. We checked that stability is independent of the amplitude of the noise provided that
it is much smaller than the amplitude of the steady state.

It is important to note that
while in all three cases the population of low-momentum states is much larger than that below threshold,
there is no long-range range order in the final states in Figs.~\ref{fig:stab-noise}(b) and~\ref{fig:stab-noise}(c), 
even in the mean field limit. 
As shown in Fig.~\ref{fig:stab-noise}(d), the coherence length
is of the order of the typical size of the structures visible on the figures. The reason for the reduced coherence
is different than in the case of 1D quasicondensates of ultracold atoms~\cite{Petrov2000,Bagnato1991,Bouchoule2007}
since here we work in the mean-field limit, and the
same instability is present also in the higher-dimensional versions of the model.
The above modulational instability of the condensate was first reported in~\cite{wouters2007}, where it was
named the ``hole-burning effect''.
The analytical condition for stability was derived in \cite{ostrovskaya2014}
\begin{equation}
\label{condition}
\frac{P}{P_{th}}>\frac{g_R}{g}\frac{\gamma}{\gamma_R}.
\end{equation}
Stability limit for $g_R=2g$ is marked in Fig.~\ref{fig:with-p}(a) with a thick black line. 

The above result has important
practical consequences. As shown in several experiments, the lifetime of polaritons is typically much shorter than the 
exciton lifetime. This can be easilly understood since the cavity photon lifetime even 
in very high-Q cavities~\cite{Snoke_BallisticMotion} 
is still shorter than the natural lifetime of excitons, which are of the order of hundreds of picoseconds.
The reported exciton lifetimes were as high as $\gamma_R^{-1}=700$ ps~\cite{Szczytko_ExcitonFormation,Bloch_DynamicsElectronGas} or
$\gamma_R^{-1}=300$ ps~\cite{Bloch_1DAmplification}. Taking into account that typical polariton lifetimes $\gamma^{-1}$ 
are of the order of a few or a few tens of picoseconds,
these values correspond to values of $\gamma/\gamma_R$ in the phase
diagram Fig.~\ref{fig:with-p}(a) where stability can be achieved only at a very high pumping powers. 

%This is in apparent
%contradiction with a number of other experiments, in which high degree of coherence 
%in the whole condensate area was demonstrated~\cite{Kasprzak_BEC,Yamamoto_SpatialCoherence,Bloch_ExtendedCondensates,Hofling_1DcoherenceInteractions}. 
Here, we suggest how this fact can be reconciled with the emergence of large coherence length in experiments. 
It was pointed out that the lifetime of active excitons, 
that is excitons that can scatter directly to the condensate,
can be much shorter than the average lifetime of excitons in the system~\cite{Deveaud_VortexDynamics}. 
Since these excitons become dressed with light field at low momenta, their lifetime can be reduced
due to a nonzero photonic Hopfield coefficient.
However, there is no fundamental reason why the lifetime should become shorter than the photon lifetime,
which is necessary for stability at {\em any power} $P>P_{th}$ in Eq.~(\ref{condition}). 
On the other hand, we show that this condition can be relaxed
in the case of a finite pumping spot. We show that in the case of inhomogeneous pumping profiles
large coherence lengths can emerge  
for relatively large values of $\gamma/\gamma_R$.

We note that while the above calculations take into account only the classical fluctuations in the initial polariton 
field, it is possible to include the effect of quantum fluctuations 
on the level of classical fields approximation~\cite{Ciuti2005,WoutersSavona2009,Matuszewski2014}. 
In the truncated Wigner approximation (TWA), the quantum field is simulated by an ensemble of realizations
of the Gross-Pitaevskii equations in the form similar to~(\ref{psi})-(\ref{nr}), 
with the addition of a stochastic term. The
extended version of Eq.~(\ref{psi}) reads $d\psi= (\dots) dt + dW$, where
$dW$ is a complex stochatic variable with~\cite{Ciuti2005,WoutersSavona2009,Matuszewski2014}
\begin{align} \label{dW}
\langle dW(x) dW(x')\rangle &= 0\,, \\ \nonumber
\langle dW(x) dW^*(x')\rangle &= \beta\frac{dt}{2 \Delta x} (R n_R + \gamma_C) \delta_{x,x'}\,,
\end{align}
reflecting the quantum noise due to the particles entering and leaving the condensate.
In Fig.~\ref{fig:stab-noise} (bottom panel, dashed lines) we show the effect of quantum fluctuations on the correlation functions.
While quantum fluctuations lead to decay of correlation 
functions over long distance~\cite{Chiocchetta2013,Gladilin2013,Bloch_ExtendedCondensates}, this
effect is overwhelmed by the instability in the unstable case (c) (black lines), which marks a dramatic reduction of the correlation length.

%% \bibitem{ClassicalFields}
%% I. Carusotto and C. Ciuti, Phys. Rev. B {\bf 72}, 125335 (2005);
%% M. Wouters and V. Savona, Phys. Rev. B {\bf 79}, 165302 (2009);
%% M. Matuszewski and E. Witkowska, Phys. Rev. B 89, 155318 (2014).

%% \bibitem{1D_Quasicondensates}
%% A. Chiocchetta and I. Carusotto, Europhys. Lett. {\bf 102}, 
%% V. Gladilin, K. Ji, and M. Wouters, arXiv:1312.0452.

%===============================================================================
\subsection{Stability of a uniform flow}
%===============================================================================
Another important issue in the case of uniform pumping is the stability of a condensate with a finite momentum.
In this case the stationary solution is
\begin{equation}
\psi(x,t)=\psi_0 e^{ipx-i\mu_0 t},
\label{steady-state-momentum}
\end{equation}
where $p$ is the condensate momentum and ${\mu_0=g\psi_0^2+g_Rn_R^0+p^2}$ the chemical potential.
Such polariton flow can occur naturally eg.~in the presence of a spatially
varying exciton-photon detuning which generates a potential gradient, or due to repulsive interactions with
reservoir excitons which are generated by the pump~\cite{Bloch_ExtendedCondensates}.
Similarly to the case of a static condensate~\cite{wouters2007,ostrovskaya2014},
the modulational stability of the steady-state solution 
(\ref{steady-state-momentum}) can be investigated within the Bogoliubov-de Gennes 
approximation~\cite{Wouters_CriticalVelocity,Ostrovskaya_PersistenCurrents}.
Small fluctuations around the steady state have the form
\begin{flalign} 
%\begin{eqnarray}
%\begin{equation}
%\begin{split}
\label{bogo-uv}
\nonumber
\psi&=\psi_0 e^{ipx-i\mu_0 t}\Bigg[ 1 +\\
&+ \sum_{k} \left\lbrace a_ke^{-i(\omega_k t-kx)}+b_k ^* e^{i(\omega_k ^*t-kx)}\right\rbrace\Bigg],\\
%\label{bogo-w}
n_R(x,t)&=n_R ^0\left[1
%\nonumber
+ \sum_{k} \left\lbrace c_k e^{-i(\omega_k t-kx)}+c_k ^*e^{i(\omega_k ^*t-kx)}\right\rbrace\right], \nonumber
%\end{split}
%\end{equation}
\end{flalign}
where $\omega_k$ is the frequency of the mode with the wavenumber $k$, and $a_k,b_k,c_k$ are small fluctuations. Substituting Eqs.~(\ref{bogo-uv}) into the system of Eqs.~(\ref{psi}) and keeping linear terms only, we get the standard eigenvalue problem $\mathcal{L}_k \mathcal{U}_k=\omega_k \mathcal{U}_k$, where ${\mathcal{U}_k=(a_k, b_k, c_k)^T}$ and
\begin{align}
&\mathcal{L}_k =\\ \nonumber
&= \begin{pmatrix}
  g\psi_0^2+2kp+k^2 & g{\psi_0}^2 & \frac{i}{2}R+g_R\\
  -g\psi_0^2 & -g\psi_0^2+2kp-k^2 &  \frac{i}{2}R-g_R \\
%  \vdots  & \vdots  & \ddots & \vdots  \\
 -i\gamma \psi_0^2 & -i\gamma \psi_0^2  & -i\left(\gamma_R+R\psi_0^2\right)
 \end{pmatrix}
 \label{matrix1}
\end{align}
The existence of an eigenvalue $\omega_k$ with a positive imaginary part marks the  
dynamical instability of the flow. 

We solved the above eigenvalue problem numerically to obtain the stability limits in parameter space.
The results are presented in Fig.~\ref{fig:with-p}, where the stability of the flow with a finite momentum are
marked with a dashed line. Similarly to~\cite{Ostrovskaya_PersistenCurrents}, the region
of stability of solutions is further decreased with respect to the static case. There is a certain area for small $P/P_{th}$ and 
$\gamma/\gamma_R$ in which the condensate becomes unstable. 
In this region, the dynamics of the instability is qualitatively different
to the one presented in Fig.~\ref{fig:stab-noise}. Instead of chaotic dynamics, regular waves are created by the unstable flow,
as shown in Fig.~\ref{fig:with-p}(b). Here, instead of periodic boundary conditions, we used boundary conditions that
provide a flow of particles with a prescribed momentum through the borders of the computational window: $\partial \psi/\partial x=ip\psi$.

The origin of the above modulational instability of the flow is clearly due to the movement of polaritons with respect to the 
practically immobile reservoir excitons. This instability is absent in the standard version of the polariton model which does not include
a separate equation for the reservoir~\cite{Berloff_VortexLattices,Wouters_CriticalVelocity},
since in this case the system is Galilean invariant. However, it can be recovered by inclusion of frequency dependent
pumping~\cite{Wouters_CriticalVelocity}, which resembles momentum dependence of scattering from the immobile reservoir.
Notably, similar to~\cite{Wouters_CriticalVelocity}, the modulational instability does not correspond to $p$ being higher than
the critical velocity, see dash-dotted line in Fig.~\ref{fig:with-p}(a). 
The latter conditon defines the limit on stability of the flow with respect to scattering against defects.

%--------------------------------------------------------------
%\subsection{Critical velocity condition}
%--------------------------------------------------------------
%According to \cite{Wouters_CriticalVelocity}, the expression for critical velocity $v_c\approx\sqrt{g^{\rm 1D}_C|\psi|^2/m^*}$ after rescaling takes the dimensionless form $\widetilde{v_c}\approx\sqrt{2\widetilde{g}|\widetilde{\psi}|^2}$, where for stationary state and homogeneous pumping ($P=const$) the condensate density is taken as ${\widetilde{\psi^2}=(P/\widetilde{\gamma})-(\widetilde{\gamma_R}/\widetilde{R})}$. The velocity of polaritons is $v'=2p$ ($m=1/2$). The fulfilment of condition $v_c<v'$ leads to condensate instability.
% can predict the instability limit}.

%--------------------------------------------------------------

\section{Effects of polariton energy relaxation}

\begin{figure}[tbp]
	\begin{center}
		\includegraphics[width=0.89\columnwidth]{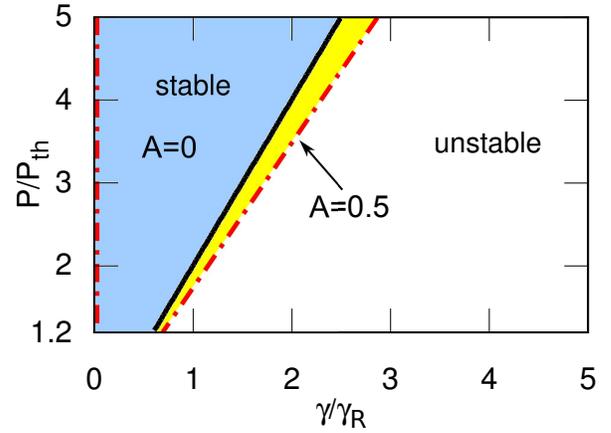} 
		\caption{(Colour online).~Comparison of stability regions with and without the energy relaxation term included. In both cases $p=0$. Red dash-dot lines mark out the stable  region for $A=0.5$, which partially overlaps with the stable region for $A=0$. Parameters are ${R=1}$, ${g=1}$, ${g_R=2g}$, ${\gamma_R=1}$, $\gamma\in[0,5]$.}
		\label{fig:with-A}
	\end{center}
\end{figure}
%--------------------------------------------------------------

One of the drawbacks of the Gross-Pitaevskii model based on Eqs.~(\ref{psi}) is that the relaxation of polariton energy
is not taken into account. 
Several experiments~\cite{Bloch_ExtendedCondensates,Bloch_1DAmplification,Bloch_GapStates} have shown that  
energy relaxation (or thermalization) may play an important role in the dynamics. 
It is important to investigate whether it can have a stabilizing effect on the condensate.
To estimate the effect of relaxation we follow Refs.~\cite{Bloch_1DAmplification,Bloch_GapStates} 
by adding a phenomenological relaxation term to the condensate evolution equation~(\ref{psi}) 
\begin{eqnarray}
\label{eq:bloch}
\frac{\partial \psi}{\partial t}&=&\\ \nonumber
&=&\left[ \left(i+A\right)\frac{\partial^2}{{\partial x}^2}+\frac{1}{2}\left(Rn_R-\gamma\right)-ig|\psi|^2-ig_Rn_R\right]\psi,
\end{eqnarray}
where the real coefficient $A$ corresponds to the energy relaxation in the condensate.
Following the steps of the previous section, we obtain the matrix which describes the Bogoliubov-de Gennes excitations
in the static ($p=0$) case
\begin{align}
&\mathcal{L}_k = \\ \nonumber
&= \begin{pmatrix}
  g\psi_0^2+i\alpha k^2 & g{\psi_0}^2 & \frac{i}{2}R+g_R\\
  -g\psi_0^2 & -g\psi_0^2+i \alpha k^2 &  \frac{i}{2}R-g_R \\
%  \vdots  & \vdots  & \ddots & \vdots  \\
 -i\gamma \psi_0^2 & -i\gamma \psi_0^2  & -i\left(\gamma_R+R\psi_0^2\right)
 \end{pmatrix}
 \label{matrix1}
\end{align}
where $\alpha=i+A$. 

The correction to the stability diagram with the relaxation term included is shown in 
Fig.~\ref{fig:with-A}, where we compare the stability limits obtained numerically by solving~(\ref{matrix1}) with
the analytical condition for the relaxation-free case~(\ref{condition}). 
It is clear that even the large relaxation rate ($A=0.5$) cannot lead to the 
stabilization of the condensate in a large parameter range, although the region of stability is increased
with respect to the $A=0$ case.

We note that excitation spectrum of a formally similar hybrid Boltzmann-Gross Pitaevskii model 
was calculated recently~\cite{Malpuech_Hybrid}. In this model, steady states were found to be dynamically (modulationally) stable 
in the whole parameter range. This stability is a consequence of the assumption
of an ideally thermalized reservoir, which translates into immediate response of the 
reservoir to the change of the condensate density. This includes the response of the reservoir density distribution $n_R(x,t)$. 
To describe the instability shown in Fig.~\ref{fig:stab-noise}, 
it is important to take into account that the relatively heavy excitons remain practically immobile on a
short time scale, so that their density distribution can become out of equilibrium 
(although the reservoir may be thermalized locally).

%--------------------------------------------------------------

%===============================================================================
\section{Inhomogeneous pumping}
%===============================================================================

\begin{figure}[tbp]
	\begin{center}
	\includegraphics[width=1.\columnwidth]{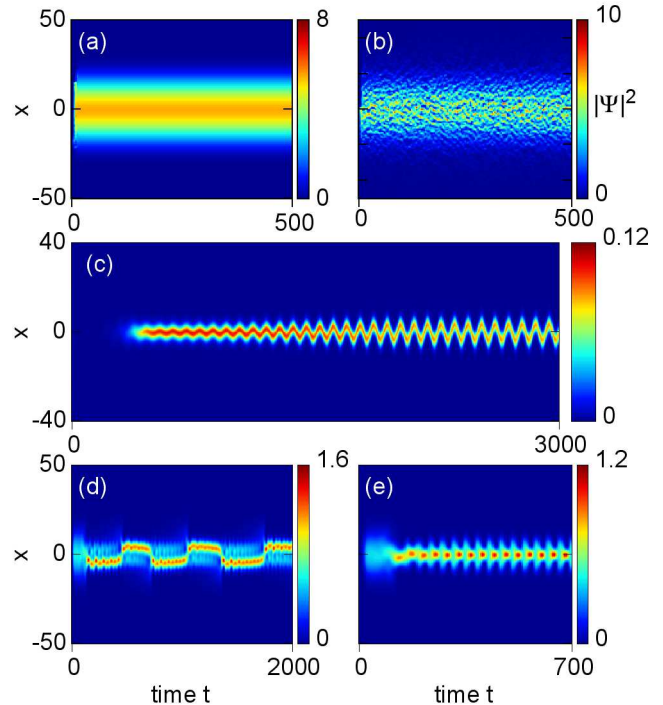} 
		\caption{(Colour online).~Evolution of the polariton density for Gaussian pumping ${P(x)=P_{\rm max} \exp(-{x^2}/{W^2})}$ 
($A=0$). Frames (a) and (b)  show condensate densities in the coherent and incoherent cases, respectively. Parameters are $R=1$, $\gamma/\gamma_R=1.5, 5 $ for (a) and (b), respectively, $g=0.38$, $P_{max}/P_{th}=1.7$, $W=44.7$.
Frames below show examples of coherent nonstationary states: (c), (d) oscillating wavepackets with parameters $R=2.5$, $\gamma/\gamma_R=0.15$, $g=3$ and $R=1$, $\gamma/\gamma_R=0.7$, $g=0.38$,  $P_{max}/P_{th}=1.8$, $W=15.8$, respectively, and (e) a breather with parameters $R=1$, $\gamma/\gamma_R=0.7$, $g=0.38$, $P_{max}/P_{th}=1.6$, $W=15.8$.}
		\label{fig:gauss-cos}
	\end{center}
\end{figure}

\begin{figure}[tbp]
	\begin{center}
		\includegraphics[width=0.89\columnwidth]{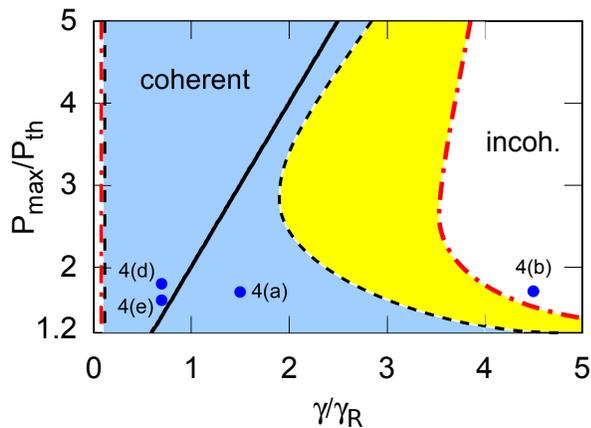} %depending on the latex compiler, you can omit the file extension
		\caption{(Colour online).~Stability regions for Gaussian pumping profiles. 
The lines mark out the coherent regions for narrower pumping of width $W=15.8$ (dash-dotted lines)
and for wider pumping of width $W=44.7$ (dashed lines). Black solid line corresponds to the analytical stability condition of a homogeneous 
condensate~\cite{ostrovskaya2014}. Blue points correspond to particular cases from Fig.~\ref{fig:gauss-cos}. Parameters are ${R=1}$, ${g=0.38}$, ${g_R=2g}$, and ${\gamma_R=1}$. For the choice of scaling parameters as in Fig.~\ref{fig:stab-noise}, the width of the pumping is
$W=54\,\mu$m (dash-dotted lines) and $W=152\,\mu$m (dashed lines).}
		\label{fig:with-gauss}
	\end{center}
\end{figure}

Experiments with nonresonantly pumped polariton condensates are performed using inhomogeneous, typically Gaussian-shaped,
pumping beams. In this section we investigate how the shape of the pump can influence the stability
of the condensate. 

In Fig.~\ref{fig:gauss-cos} we show examples of dynamics for various parameters of the system with a Gaussian pumping profile,
${P(x)=P_{\rm max} \exp(-{x^2}/{W^2})}$.  The initial state is a small white noise in the polariton field $\psi(x)$. 
%{\color{red} what about dark solitons?}
The frames (a) and (b) show examples of stable and unstable dynamics, corresponding to Fig.~\ref{fig:stab-noise}(a) and (c).
There is also a number of other possible nonstationary states, as shown in Fig.~\ref{fig:gauss-cos}(c)-(e), including the oscillating
wavepackets and breathers. These states, despite the complex dynamics, display correlation length that 
is comparable with the size of the Gaussian pump.
For this reason, we classify these states as ``coherent'' in contrast to the ``incoherent'' unstable state of Fig.~\ref{fig:gauss-cos}(b).
We note that nonstationary  condensate states have been previously shown to emerge in two-dimensional models with inhomogeneous 
pumping~\cite{Berloff_VortexLattices,Berloff_PatternFormation}.

In Fig.~\ref{fig:with-gauss} we show the regions of parameter space where coherent (stationary or nonstationary) states exist. The limits of the regions were determined numerically by solving the Gross-Pitaevskii equation~(\ref{psi}).
Clearly, the region of coherence in the case of inhmogeneous pumping is much larger 
than in the homogeneous case of~Fig.~\ref{fig:with-p}. Moreover, it expands significantly as the pump size is decreased. 
The reduction of the pumping spot appears to be a very effective way to stabilize
condensates even at high ratios of $\gamma/\gamma_R$. 
Condensates with $\gamma/\gamma_R = 3.5$ are fully coherent at all pumping powers already at 
a relatively wide pumping spot with $W=15.8$. 
We note that the threshold value of $P_{max}$ for the appearence of 
the condensate in the case of inhomogeneous pumping is higher than $P_{th}$~\cite{Ostrovskaya_DissipativeSolitons}.
However, since our pumping profiles are relatively wide, this difference cannot be seen in the 
area covered by Fig.~\ref{fig:with-gauss} which begins at $P/P_{th}=1.2$.

\section{conclusions}

In conclusion, we investigated the dynamical stability and its relation to 
spatial coherence properties of one-dimensional exciton-polariton condensates under
nonresonant pumping. In the case of spatially homogeneous pumping, we found that the instability of the steady state
leads to a significant reduction of the condensate coherence. 
This instability is predicted to occur in the physically relevant case when the loss rate of the exciton reservoir is 
lower than the loss rate of the exciton-polaritons. 
Since the experiments with inhomogeneous optical pumping reported large coherence 
lengths~\cite{Kasprzak_BEC,Yamamoto_SpatialCoherence,Bloch_ExtendedCondensates,Hofling_1DcoherenceInteractions,Yamamoto_RMP},
we considered two effects that can potentially lead to the stabilization
of condensates. These are the effect of polariton energy relaxation and the inhomogeneous pumping profiles.
We found that while the former has little effect on the stability, the latter is very effectve in stabilizing the condensate
which results in a large coherence length.

It is noteworthy that in the two-dimensional experiment with a top-hat shaped pumping profile, which is the closest 
to the homogeneous profile, strong internal dynamics and vortex creation 
were observed~\cite{Yamamoto_VortexPair}, which evidences the existence of a nonstationary state. However,
we believe that further experiments, especially with almost-homogeneous or ring-shaped pumps, 
are necessary to determine the stability limits in the parameter
space. Such experiments would also provide verification of the open-dissipative Gross-Pitaevskii model~(\ref{GP1}) and
its modifications
widely used in the literature, and allow for the determination
of their phenomenological parameters.

\acknowledgments

N.B.~and M. M.~acknowledge support from the National Science Center grant DEC-2011/01/D/ST3/00482. E.A.O.~acknowledges support by the Australian Research Council (ARC).

\bibliography{references}

\end{document}